\documentclass[12pt]{article}
\usepackage{amssymb,amsmath,cite,geometry,graphicx,url,fancybox,boxedminipage,curves,mathrsfs}
\catcode`\@=11

\@addtoreset{equation}{section}
\def\theequation{\arabic{section}.\arabic{equation}}
     
\catcode`\@=11
\def\thesection{\arabic{section}.}

\def\appendix{\setcounter{section}{0}
        \def\thesection{Appendix.}
        \def\theequation{\Alph{section}.\arabic{equation}}}
\def\section{\@startsection{section}{1}{\z@}{3.5ex plus 1ex minus
   .2ex}{2.3ex plus .2ex}{\large\bf}}
      
\long\def\@makefntext#1{\parindent 0cm\noindent
\hbox to 1em{\hss$^{\@thefnmark}$}#1}

\newcommand{\captionfonts}{\small}
\makeatletter  
\long\def\@makecaption#1#2{%
  \vskip\abovecaptionskip
  \sbox\@tempboxa{{\captionfonts #1: #2}}%
  \ifdim \wd\@tempboxa >\hsize
    {\captionfonts #1: #2\par}
  \else
    \hbox to\hsize{\hfil\box\@tempboxa\hfil}%
  \fi
  \vskip\belowcaptionskip}
\makeatother   
 
\begin{document}
\begin{titlepage}
\vspace{.5in}
\begin{flushright}
August 2016\\[.2ex] 
revised July 2017\\ 
\end{flushright}
\vspace{.5in}
\begin{center}
{\Large\bf
 The Dynamics of Supertranslations\\[1ex]  and Superrotations in 2+1 Dimensions}\\  
\vspace{.4in}
{S.~C{\sc arlip}\footnote{\it email: carlip@physics.ucdavis.edu}\\
       {\small\it Department of Physics}\\
       {\small\it University of California}\\
       {\small\it Davis, CA 95616}\\{\small\it USA}}
\end{center}

\vspace{.5in}
\begin{center}
{\large\bf Abstract}
\end{center}
\begin{center}
\begin{minipage}{4.75in}
{\small
Supertranslations, and at least in 2+1 dimensions superrotations, 
are asymptotic symmetries of the metric in asymptotically flat
spacetimes.  They are not, however, symmetries of the boundary
term of the Einstein-Hilbert action, which therefore induces an 
action for the Goldstone-like fields that parametrize these symmetries.  
I show that in 2+1 dimensions, this action is closely related to a
chiral Liouville action, as well as the ``Schwarzian'' action that appears 
in two-dimensional near-AdS physics.
}
\end{minipage}
\end{center}
\end{titlepage}
\addtocounter{footnote}{-1}

Asymptotically flat spacetime looks asymptotically like Minkowski space.  
One might therefore expect its asymptotic symmetries to be the
symmetries of Minkowski space, the Poincar{\'e} group.  Surprisingly,
this is not the case at null infinity $\mathscr{I}^\pm$: it has been understood
since the 1960s \cite{BMS1,BMS2} that the symmetries are described by
the larger BMS group, which includes angle-dependent supertranslations.  
These symmetries have recently received renewed attention, thanks in part 
to their relationship to soft graviton theorems \cite{Stroma,Stromaa}, the 
gravitational memory effect \cite{Stromb}, and perhaps the black hole 
information loss problem \cite{HPS}.

The goal of this paper is to show that the Goldstone-like excitations associated
with these symmetries acquire a dynamics, induced from the boundary term
in the action at $\mathscr{I}^\pm$.  For simplicity, I focus here on the case
of (2+1)-dimensional spacetimes, where the BMS group also includes 
superrotations \cite{AshB}.\footnote{The role of superrotations in (3+1)-%
dimensional spacetimes is not yet clear \cite{BarnTroa,CompLong}.}  The 
resulting boundary action includes a chiral Liouville theory, which is also closely 
related to the ``Schwarzian action'' that appears in 
nearly anti-de Sitter gravity in two dimensions \cite{Jensen,Eng,Mal}, and 
it has intriguing connections to the coadjoint orbit quantization of the Virasoro 
group \cite{AlexShat,Witten,Balog,SJ}.  Chiral Liouville theory has previously 
been associated with BMS$_3$ \cite{BarnGomberoff,BarnGonzalez}, but in a 
somewhat less direct way, via a Chern-Simons formulation whose generalization 
to higher dimensions seems problematic.

The basic idea is fairly simple.  If one starts with a theory with an action
 $I_{\hbox{\tiny\it bulk}}$ and places it on a manifold with boundary, one must
 usually add a boundary term $I_{\hbox{\tiny\it bdry}}$ to the action.  Classically,
 $I_{\hbox{\tiny\it bdry}}$ is required for the existence of extrema: a variation
 of the bulk action gives boundary terms from integration by parts that must be
 cancelled off.  Quantum mechanically, $I_{\hbox{\tiny\it bdry}}$ is required for
 the proper ``sewing'' of path integrals, the path integral analog of a sum over
 intermediate states.  The key observation \cite{Carlip1} is that even if the
 bulk action is gauge invariant, the boundary action may not be.  Hence
 field configurations that would normally be considered physically equivalent
 can become distinct at the boundary, giving rise to new degrees of freedom.
 The action for these ``would-be gauge degrees of freedom'' is induced from
 $I_{\hbox{\tiny\it bdry}}$, and can sometimes be calculated explicitly.  In
 particular, for (2+1)-dimensional asymptotically anti-de Sitter gravity, it is
 a Liouville theory \cite{CS,CS2,Carlip2} with the proper central charge to match
 the Brown-Henneaux asymptotic symmetry \cite{BrownHenn}.  We shall see
 here that a similar conclusion holds for the asymptotically flat case.
 
\section{Metric and diffeomorphisms}

\begin{figure}
\scalebox{.75}{
\begin{picture}(300,180)(-180,100)
\thicklines
\put(0,200){\line(1,1){100}}
\put(0,200){\line(1,-1){100}}
\put(200,200){\line(-1,1){100}}
\put(200,200){\line(-1,-1){100}}
\put(100,210){\line(1,1){45}}
\put(100,210){\line(-1,1){45}}
\curve(56,255,100,245,144,255)
\curve(0,200,100,185,200,200)
\curvesymbol{\phantom{\circle*{2}}\circle*{2}}
\curve[-10](56,255,100,262,144,255)
\curve[-20](0,200,100,218,200,200)
\put(142,270){$\scriptstyle\mathscr{I}^+$}
\put(142,130){$\scriptstyle\mathscr{I}^-$}
\put(98,304){$i^+$}
\put(204,197){$i^0$}
\put(138,235){\vector(-2,1){6}}
\thinlines
\curve(137,235,170,230,198,230)
\put(202,227){\tiny $u$\,=\,\it{const.}}
\end{picture}}
\label{fig1}
\caption{A (2+1)-dimensional spacetime foliated by null cones $u=\hbox{const.}$}
\end{figure}
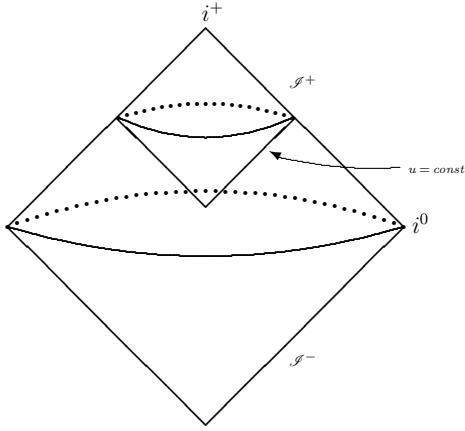
To discuss the behavior of the metric near (future) null infinity, it is useful to
choose coordinates in which the approach to $\mathscr{I}^+$ is easy to describe.
Bondi coordinates \cite{BMS1} are defined by requiring that near infinity, spacetime 
is foliated by outgoing null cones, here labeled  by a coordinate $u$ (see figure 1).
In 2+1 dimensions, the metric then takes the form
\begin{align}
ds^2 =  -2V dudr + g_{uu}\, du^2 + 2g_{u\phi}\, dud\phi + r^2e^{2\omega} d\phi^2
\label{a1}
\end{align}
where
\begin{align}
V=\mathcal{O}(1), \quad g_{uu} = \mathcal{O}(r), \quad g_{u\phi} = \mathcal{O}(1), \quad
\omega = \omega_0 + \frac{\omega_1}{r} + \dots
\label{a1a}
\end{align}
With the additional choice gauge $V=1$, the radial coordinate $r$ is an affine parameter
for the geodesics generating the null cones.  For convenience, I will make this choice
here; it does not affect the final conclusions.

Barnich and Troessaert have analyzed the vacuum field equations for such a metric  
\cite{BarnTrob}.  With the additional restriction  $\omega_1=0$, they find that
\begin{align}
&g_{uu} \sim -2r\partial_u\omega + e^{-2\omega}\left[ -(\partial_\phi\omega)^2 
   + 2 \partial_\phi^2\omega + \Theta\right] \nonumber\\
&g_{u\phi} \sim e^{-\omega}\left[\Xi + \int^u d{\tilde u} \left\{ \frac{1}{2}\partial_\phi\Theta
  - \partial_\phi\omega\left[\Theta - (\partial_\phi\omega)^2 + 3\partial_\phi^2\omega\right] 
  + \partial_\phi^3\omega\right\}\right]\nonumber\\[1ex]
&\hbox{with $\partial_u\Theta = \partial_u\Xi =0$}
\label{a2}
\end{align}
near $\mathscr{I}^+$.
The metric thus depends on the conformal factor $\omega$ and two functions $\Theta$ 
and $\Xi$ of the angular coordinate $\phi$, which can be shown to be the charges 
associated with supertranslations and superrotations.

We next evaluate the action of diffeomorphisms on this metric.  Let us start with
the standard flat metric
\begin{align}
ds^2 = -2d{\bar u}d{\bar r}- d{\bar u}^2 + {\bar r}^2d{\bar\phi}^2
\label{a3}
\end{align}
and consider the diffeomorphisms
\begin{align}
{\bar u} = u_0 + \frac{u_1}{r} + \dots, \quad
{\bar\phi} = \phi_0 + \frac{\phi_1}{r} + \frac{\phi_2}{r^2} +\dots, \quad
{\bar r} = ar + b_0 + \dots
\label{a4}
\end{align}
where the coefficients are functions of new coordinates $\phi$ and $u$.  For
transformations that preserve the asymptotic form (\ref{a1}) of the metric, the
functions $\phi_0$ and $u_0$---the asymptotic reparametrizations of the circle and 
$\phi$-dependent translations of $u$---are the superrotations and supertranslations. 

The requirement that the new metric be of the form (\ref{a1}) with $V=1$ and 
$\omega_1=0$ leads, after a straightforward calculation, to the conditions%
\footnote{I show in the appendix that the conclusions are essentially unchanged 
if $\omega_1\ne0$.}  
\begin{align}
&a\partial_u u_0 = 1,\quad \partial_u\phi_0 =0,\quad
  e^{-\omega_0}a\partial_\phi\left(\frac{\partial_\phi u_0}{\partial_\phi \phi_0}\right) = b_0 
  \nonumber\\[.3ex]
&u_1 = -\frac{a}{2}\phi_1{}^2,\quad \phi_1 = -\frac{1}{a}\frac{\partial_\phi u_0}{\partial_\phi\phi_0},\nonumber\\[.3ex]
&\phi_2 = -\frac{b_0}{a}\phi_1 
\label{a5}
\end{align}
The components of the metric are then
\begin{align}
&e^{\omega_0} = \frac{\partial_\phi\phi_0}{\partial_u u_0} , \quad
  \nonumber\\[.5ex]
&g_{uu} = -2r\partial_u\omega_0 
  + e^{-2\omega_0}\left[  - (\partial_\phi\omega_0)^2 + 2\partial_\phi{}^2\omega_0
  -(\partial_\phi\phi_0)^2 -2\{\phi_0;\phi\}\right] \nonumber\\[.5ex]
&g_{u\phi} = -e^{-\omega_0}\left[
  \partial_\phi{}^2\left(\frac{\partial_\phi u_0}{\partial_\phi\phi_0}\right) 
   - \frac{\partial_\phi{}^2\phi_0}{\partial_\phi\phi_0}\partial_\phi 
  \left(\frac{\partial_\phi u_0}{\partial_\phi\phi_0}\right)
  + (\partial_\phi\phi_0)^2\left(\frac{\partial_\phi u_0}{\partial_\phi\phi_0}\right)^2\right]
\label{a6}
\end{align}
where the Schwarzian derivative $\{\phi_0;\phi\}$ in $g_{uu}$ is defined by
\begin{align}
\{f;z\} =  \frac{f^{\prime\prime\prime}}{f'} - \frac{3}{2}\left(\frac{f^{\prime\prime}}{f'}\right)^2
\label{a7}
\end{align}
It is not hard to check that these results match (\ref{a2}), but with coefficients that
are now explicit functions of the parameters that label superrotations and supertranslations.
In particular,
\begin{align}
\Theta = -(\partial_\phi\phi_0)^2 -2\{\phi_0;\phi\}
\label{a8}
\end{align}
($\Xi$ becomes a complicated function of the $u$-independent part of 
$\partial_\phi u_0/\partial_\phi\phi_0$; we will not need its explicit form.)

\section{Boundary terms, corner terms, and the induced action}

To proceed further, we shall need the boundary term for the action.  For a spacelike or timelike
boundary with a fixed induced metric, the proper choice is the Gibbons-Hawking term 
\cite{Gibbons}, an integral of the extrinsic curvature of the boundary.  For a lightlike boundary 
like $\mathscr{I}^\pm$, the choice is less clear: the null normal to the boundary has no 
preferred normalization, so the analog of the extrinsic curvature, the expansion, is not
unique.  I will therefore take the less elegant approach of directly computing the boundary 
terms in the variation of the action and finding a boundary action to cancel them.

We start with the standard Einstein-Hilbert action 
\begin{align}
I = \frac{1}{\kappa^2}\int_M \!d^3x\,\sqrt{-g}R \qquad (\hbox{with}\ \kappa^2=16\pi G_N)
\label{b1}
\end{align} on a manifold $M$ with metric (\ref{a1}) and a boundary at $r={\bar r}$; we will
take the limit ${\bar r}\rightarrow\infty$ at the end.  A standard calculation gives
\begin{align}
\delta I &= \hbox{\it e.o.m.} + \frac{1}{\kappa^2}\int_{r={\bar r}}\!d^2x\,\sqrt{-g}
  \left[g^{ab}\delta\Gamma^r_{ab} - g^{ar}\delta\Gamma^b_{ab}\right] \nonumber\\
&= \dots + \frac{1}{\kappa^2} \int_{r={\bar r}}\!d^2x\,\sqrt{-g}g^{ab}g^{rc}
  \left[\nabla_a\delta g_{bc} - \nabla_c\delta g_{ab}\right]
\nonumber\\
&= \dots - \frac{1}{\kappa^2}\int_{r={\bar r}}\!d^2x\, 
  \left[\partial_a(\sqrt{-g}\,\delta g^{ra}) + \sqrt{-g}\,\Gamma^r_{ab}\delta g^{ab}
  - \sqrt{-g}\, g^{rc}\partial_c(g_{ab}\delta g^{ab})\right]
\label{b2}
\end{align}
As in \cite{BarnTrob}, let us set $V=1$ and $\omega_1=0$ in (\ref{a1}).  A short computation 
then gives
\begin{align}
\delta I = \dots + \frac{1}{\kappa^2}\int_{r={\bar r}}\!d^2x\,  \left[ 2r\partial_u(e^\omega\delta\omega)
  + \partial_r(re^\omega\delta g_{uu}) + 2g_{uu}\partial_r(re^\omega)\delta\omega \right]
  + \mathcal{O}\left({\bar r}^{-1}\right)
\label{b3}
\end{align}
Using the asymptotic form (\ref{a2}) of the metric, this expression simplifies to
\begin{align}
\delta I = \dots + \frac{1}{\kappa^2}\int_{r={\bar r}}\!d^2x\,  \left[ 
   -2r\partial_u(e^{\omega_0}\delta\omega_0)
   + \delta(e^{\omega_0}g_{uu}^{(0)})
  +  g_{uu}^{(0)}\delta(e^{\omega_0})\right] + \mathcal{O}\left({\bar r}^{-1}\right)
\label{b4}
\end{align}
where $g_{uu}^{(0)}$ means the $\mathcal{O}(1)$ part of $g_{uu}$.   

Now, (\ref{b4}) is not, in general, a variation of any boundary action.  This is to be expected:
for a variational principle to make sense, we need to impose boundary conditions on some
of the phase space variables.  In fact, (\ref{b4}) tells us, roughly, that $g_{uu}$ and $e^\omega$
are canonically conjugate variables, a fact that can be confirmed by looking at radial evolution
in the Hamiltonian formalism.  To choose boundary conditions, note that the conformally
compactified metric near $\mathscr{I}^+$ is
\begin{align}
d{\bar s}^2 = \rho^2ds^2 \sim 2dud\rho + e^{2\omega}d\phi^2
\label{b5}
\end{align}
were $\rho = 1/r$.  The only non-gauge-fixed component of the metric at $\mathscr{I}^+$ 
is $e^\omega$, so it makes sense to hold this quantity fixed.  If $\delta\omega_0=0$
at $\mathscr{I}^+$, the boundary variation (\ref{b4}) will be cancelled by the variation of
a boundary action
\begin{align}
I_{\hbox{\tiny\it bdry}} = -\frac{1}{\kappa^2}\int_{\mathscr{I}^+}\!d^2x\,
   e^{\omega_0}g_{uu}^{(0)}  
\label{b6}
\end{align}
where $\omega_0$ is now an arbitrary but fixed function of $u$ and $\phi$.  

At finite $\bar r$, the action (\ref{b6}) is not quite the Gibbons-Hawking term:
that term is obtained by fixing the full boundary metric, while we are only fixing
$g_{\phi\phi}$.  But (\ref{b6}) can be written in an invariant form resembling the
Gibbons-Hawking term.  Let $n_a$ be the unit normal to the surface
$r={\bar r}$, and let $\ell_a$ be the null normal to the surfaces of constant $u$,
normalized so that $\ell_a n^a=-1$.  The projector
\begin{align}
q^{ab} = g^{ab} + \ell^an^b + \ell^bn^a + \ell^a\ell^b
\label{b7}
\end{align}
projects onto circles of constant $u$ and $r$.  It may then be checked that
the boundary action takes the geometric form
\begin{align}
I_{\hbox{\tiny\it bdry}} = -\frac{1}{\kappa^2}\int_{\mathscr{I}^+}\!d^2x\,
   \sqrt{-{}^{\hbox{\tiny (2)}}\!g}\, q^{ab}\nabla_an_b
\label{b8}
\end{align}

(In \cite{Grumiller}, Detournay et al.\ construct a boundary action for three-dimensional
Euclidean gravity, with the added restriction that $\omega_0=0$.  In Euclidean
signature, the extrinsic curvature is defined even in the limit ${\bar r}\rightarrow\infty$,
and Detournay et al.\ argue that the proper boundary action is one-half of the usual
Gibbons-Hawking term.  The factor of one-half matches (\ref{b8})---the standard
Gibbons-Hawking term has a prefactor of $2/\kappa^2$---and while (\ref{b8}) is not
identical to the boundary action of \cite{Grumiller}, the differences vanish when
$\omega_0=0$.)

Let us now restrict our attention to metrics of the form (\ref{a6}), that is, metrics
obtained at least asymptotically from the standard flat metric by superrotations 
and supertranslations.  The action (\ref{b6}) then becomes
\begin{align}
I_{\hbox{\tiny\it bdry}} = \frac{1}{\kappa^2}\int_{\mathscr{I}^+}\!d^2x\, e^{-\omega_0} 
  \left[   (\partial_\phi\omega_0)^2 - 2\partial_\phi{}^2\omega_0
  + (\partial_\phi\phi_0)^2 +2\{\phi_0;\phi\}\right]
\label{b9}
\end{align}
There is one subtlety, though.  From (\ref{a5}) and (\ref{a6}). we have
\begin{align}
e^{-\omega_0} = \partial_u F \quad \hbox{with}\ \   F =  \frac{u_0}{\partial_\phi \phi_0} 
\label{b10}
\end{align}
Since, moreover, $\partial_u\phi_0=0$, the last two terms in (\ref{b9}) are total derivatives,
which reduce to ``corner'' terms at the ends of $\mathscr{I}^+$.  We thus have
\begin{align}
I_{\hbox{\tiny\it bdry}} = \frac{1}{\kappa^2}\int_{\mathscr{I}^+}\!d^2x\, e^{-\omega_0} 
  \left[  (\partial_\phi\omega_0)^2 - 2\partial_\phi{}^2\omega_0 \right]
  +  \frac{1}{\kappa^2}\int_{\partial\mathscr{I}^+}\!\!d\phi\, F\left[ 
   (\partial_\phi\phi_0)^2 +2\{\phi_0;\phi\}\right]
\label{b11}
\end{align} 
This is our induced boundary action for the superrotations and supertranslations.

\section{Dynamics on $\mathscr{I}^+$}

Let us begin by considering the first integral in the boundary action (\ref{b11}).
If $\omega_0$ is fixed, as we required to obtain our boundary action, this term is 
merely a fixed constant.  This is as it should be.  The boundary term was chosen so
that the action as a whole had no boundary variation.  For a variation that is a pure
diffeomorphism,  the bulk action  (\ref{b1}) is already invariant, so the boundary 
term should be as well.

The interesting boundary dynamics comes when one allows $\omega_0$ to vary.
Different choices of $\omega_0$ correspond to different vacua, and diffeomorphisms
that change $\omega_0$ are analogous to Goldstone modes.\footnote{I believe the
comparison to Goldstone modes was first made by Kaloper and Terning, as cited in
\cite{Carlipz}.  These modes are closely related to the ``soft modes'' of Strominger
et al.\ \cite{Stroma}.}  At first sight, the boundary action (\ref{b11}) is not very
interesting dynamically, since it seems to involve only angular derivatives.  But
recall the $\omega_0$ itself contains time derivatives.  In terms of the function
$F$ defined in (\ref{b10}), the action along $\mathscr{I}^+$ is
\begin{align}
I_{\mathscr{I}^+} = -\frac{1}{\kappa^2}\int_{\mathscr{I}^+}\!d^2x\,
    \frac{\left(\partial_u\partial_\phi F\right)^2}{\partial_u F}
   = -\frac{1}{\kappa^2}\int_{\mathscr{I}^+}\!d^2x\ 
   \frac{(\partial_\phi\partial_u u_0)^2}{(\partial_\phi\phi_0)(\partial_u u_0)} 
   + \hbox{\it corner terms}
\label{x1}
\end{align}
(The first expression here is the ``right'' one, since $F$, rather than $u_0$ or $\phi_0$, 
is specified by boundary conditions, but the second exhibits the pattern of 
derivatives more clearly.)  Although it differs in detail, this action is similar in 
structure to the action of Alexeev and Shatashvili for the coadjoint orbits of the 
Virasoro group.  It would be interesting to see if it has a closer relationship to 
the corresponding action for coadjoint orbits of BMS$_3$ \cite{Oblak}.

\section{Liouville theory, the Schwarzian action, and Virasoro orbits}

We now turn the the second integral in (\ref{b11}),
\begin{align}
I_{\hbox{\tiny\it corner}} &= \frac{1}{\kappa^2}\int_{\partial\mathscr{I}^+}\!\!d\phi\, F\left[ 
   (\partial_\phi\phi_0)^2 +2\{\phi_0;\phi\}\right]
\label{c0}
\end{align}
This is a ``corner term,'' appearing at the boundaries of $\mathscr{I}^+$, that is, 
at spacelike and future timelike infinity $i^0$ and $i^+$.  The existence of such 
a term should not surprise us: the leading supertranslations and superrotations are 
time-independent, so their action should reduce to one at a fixed time.  (In
higher dimensions, a supertranslation can originate at a finite time from a pulse of
gravitational radiation arriving at $\mathscr{I}^+$---this is a form of the 
gravitational memory effect \cite{Stromb}---but in 2+1 dimensions there are
no gravitational waves.)

The action (\ref{c0}) is essentially identical to the ``Schwarzian action'' found by 
several authors \cite{Jensen,Eng,Mal} in the rather different setting of two-dimensional 
asymptotically nearly anti-de Sitter spacetime.  The physical interpretations differ,
but in both cases the action is related to deformations of circles (here at $i^0$ and $i^+$).
It seems likely that the results of \cite{Mal} on the quantum theory can be translated 
directly to this context.  

The corner term is also intimately related to chiral Liouville theory.  To see this, note 
that for (\ref{a4}) to be a diffeomorphism, $\phi_0$ must be a monotonic function 
of $\phi$, so we can write
\begin{align}
\partial_\phi\phi_0 = \sqrt{\frac{\mu}{4}}e^{\frac{\gamma}{2}\chi}
\label{c1}
\end{align}
where $\gamma$ and $\mu$ are constants and, from (\ref{a5}), $\partial_u\chi=0$.  
The corner term then becomes
\begin{align}
I_{\hbox{\tiny\it corner}} &= \frac{1}{\kappa^2}\int_{\partial\mathscr{I}^+}\!d\phi\, 
  F\left[ -\frac{\gamma^2}{4}(\partial_\phi\chi)^2 
  + \gamma\partial_\phi^2\chi + \frac{\mu}{4}e^{\gamma\chi}\right]  
\label{c2}
\end{align}
To put this into a standard Liouville form, we lift back to two dimensions,
\begin{align}
I_{\hbox{\tiny\it corner}}&= \frac{1}{\kappa^2}\int_{\mathscr{I}^+}\!d^2x\, e^{-\omega_0} 
  \left[ -\frac{\gamma^2}{4}(\partial_\phi\chi)^2 - \gamma\chi\left(
  \partial_\phi^2\omega_0 - (\partial_\phi\omega_0)^2\right)
  + \frac{\mu}{4}e^{\gamma\chi}\right]
\label{c2a}
\end{align}
and introduce an auxiliary two-dimensional metric
\begin{align}
d{\tilde s}^2 = e^{-2\omega_0} du^2 -  d\phi^2 
\label{c3}
\end{align}
with a scalar curvature 
\begin{align}
{\tilde R} = -2\left(
  \partial_\phi^2\omega_0 - (\partial_\phi\omega_0)^2\right)
\label{c4}
\end{align}
If we choose 
$\gamma^2 = \frac{\kappa^2}{2\pi}$,
the action (\ref{c2a}) becomes
\begin{align}
I_{\hbox{\tiny\it corner}} = \frac{1}{4\pi}\int_{\mathscr{I}^+}\!d^2x\, \sqrt{-{\tilde g}}
  \left[ \frac{1}{2}{\tilde g}^{ab}\partial_a\chi\partial_b\chi + \frac{1}{\gamma}\chi {\tilde R}
  + \frac{\mu}{2\gamma^2}e^{\gamma\chi}\right]
\label{c6}
\end{align}
with the restriction $\partial_u\chi=0$.  This is precisely the Liouville action
for $\chi$ \cite{Seiberg}, with a classical central charge
\begin{align}
c = \frac{12}{\gamma^2} = \frac{24\pi}{\kappa^2} = \frac{3}{2G}
\label{c7}
\end{align}
A similar chiral Liouville action was found in \cite{BarnGomberoff,BarnGonzalez},
by means of a Chern-Simons formulation, but here the meaning of the Liouville
field is clear: it is precisely the parameter that characterizes superrotations. 

To obtain this Liouville action, we incorporated $\omega_0$ into the ``background 
metric'' (\ref{c3}), implicitly treating it, and the related function $F$, as fixed quantities 
at the corners.  But it is clear from (\ref{b10}) that for a fixed $F$, the supertranslations 
and superrotations are not independent, so it should be possible to reexpress the corner 
action as an action for the supertranslation parameter $u_0$.   This is indeed the case: 
starting with (\ref{c0}) and setting
\begin{align} 
\partial_\phi\phi_0 = \frac{u_0}{F}, \quad u_0 = e ^\sigma
\label{c7a}
\end{align}
a simple calculation yields
\begin{align}
I_{\hbox{\tiny\it corner}} &= -\frac{1}{\kappa^2}\int_{\partial\mathscr{I}^+}\!d\phi\, 
   F \left[(\partial_\phi\sigma)^2 - \frac{1}{F^2}e^{2\sigma} 
   - \left(\frac{\partial_\phi F}{F}\right)^2\right]
\label{c8}
\end{align}
which again has the general form of a Liouville action.

Our boundary action also has an intriguing relationship to the quantization of the
co\-adjoint orbits of the Virasoro group \cite{AlexShat,Witten}.  It is not quite the
Alekseev-Shatashvili action of \cite{AlexShat}, which is not chiral, but it is the
integral of the chiral stress-energy tensor of that theory,
\begin{align}
T = b_0(\partial_\phi\phi_0)^2 - \frac{c}{24\pi}\{\phi_0,\phi\} \qquad
\hbox{with \ $\displaystyle b_0 = -\frac{c}{48\pi}$}
\label{c9}
\end{align}
where this value of $b_0$ corresponds to the orbit of $L_0=0$ \cite{Witten}.
This connection is further strengthened if we rewrite the boundary action
(\ref{b9}) as
\begin{align}
I_{\hbox{\tiny\it bdry}} = \frac{1}{\kappa^2}\int_{\mathscr{I}^+}\!d^2x\, e^{-\omega_0} 
  \left[   (\partial_\phi\omega_0)^2 - 2\partial_\phi{}^2\omega_0
  - \frac{48\pi}{c}T\right]
\label{c10}
\end{align}
and allow $\omega_0$ to vary while holding $T$ fixed---that is, allowing the
boundary supertranslations to vary while fixing the superrotations.  Setting 
$\psi = e^{-\omega_0/2}$, we find that the equation of motion for
$\psi$ is Hill's equation,
\begin{align}
\psi^{\prime\prime} - \frac{12\pi}{c}T\psi = 0
\label{c11}
\end{align}
where a prime means a $\phi$ derivative.  This equation has two solutions, say
$\psi_1$ and $\psi_2$.  If we set $\varepsilon = \psi_1^2$, $\psi_1\psi_2$,
or $\psi_2^2$, it is then easy to check \cite{Balog,SJ} that\footnote{I thank
Shahin Sheikh-Jabbari for explaining this point to me.}
\begin{align}
\delta_\varepsilon T = \frac{c}{24\pi}\varepsilon^{\prime\prime\prime} 
  - 2T\varepsilon' - T'\varepsilon = 0
\label{c12}
\end{align}
But $\delta_\varepsilon T$ is just the variation of $T$ under an infinitesimal
conformal transformation, that is, an action of the Virasoro group, and its
vanishing determines the coadjoint orbits of the Virasoro group.

\section{Next steps}

I have shown that the supertranslations and superrotations in asymptotically
flat (2+1)-dimensional gravity become genuine physical degrees of freedom
at null infinity.  Much as in the asymptotically anti-de Sitter case, Goldstone-like
``boundary gravitons'' are dynamical along $\mathscr{I}^+$, while additional
corner terms at spacelike and future timelike infinity induce a chiral conformal
action for the superrotations.   A good deal is known about this conformal field 
theory, though the ``nonnormalizable sector'' of Liouville theory is not fully 
understood.  It should be possible to translate the field theoretical results into 
statements about quantum gravity, although one must presumably first 
understand the identification of the past and future theories as in \cite{Stroma}.

Somewhat mysteriously, the superrotation charge $\Xi$ occurring in (\ref{a2})
does not appear in our boundary action.  This charge, which depends on the
$u$-independent part of $\partial_\phi u_0$, occurs only at subleading
order.  Further investigation is needed to see whether we are losing part
of the dynamics.  In particular, it is not clear whether the chiral Liouville
theory captures the full algebra of the BMS$_3$ group.

It would also be useful to reexpress these results more
invariantly in terms of the conformally compactified spacetime, using the
methods of \cite{AshB}.  For instance, it should be possible to express the
boundary term (\ref{b8}) in terms of quantities on the compactified
spacetime.

The main question, of course, is whether these results can be extended to a
realistic (3+1)-dimensional spacetime.  In contrast to the Chern-Simons
approach of \cite{BarnGonzalez}, the basic approach of this paper should
generalize to arbitrary dimensions, but the details may well be quite different.

\begin{appendix}

\section{Putting back $\omega_1$}

The calculations presented above have assumed that $\omega_1=0$ in eqn.\ (\ref{a1a}), as
in \cite{BarnTrob}.  Here I will describe the (minimal) changes that occur if $\omega_1$ is
allowed to be nonzero.  Note that this can be accomplished by a coordinate transformation
$r\rightarrow r+f(u,\phi)$.

The first change is that $b_0$ is no longer determined in eqn.\ (\ref{a5}).  Instead,  
\begin{align}
\omega_1 = -e^{-\omega_0}\partial_\phi\left(\frac{\partial_\phi u_0}{\partial_\phi \phi_0}\right) 
  + b_0\partial_u u_0
\label{A1}
\end{align}
This shifts the metric components in (\ref{a6}) to
\begin{align}
&g_{uu} = -2r\partial_u\omega_0 - 2e^{-\omega_0}\partial_u\left(e^{\omega_0}\omega_1\right)
  + e^{-2\omega_0}\left[  - (\partial_\phi\omega_0)^2 + 2\partial_\phi{}^2\omega_0
  -(\partial_\phi\phi_0)^2 -2\{\phi_0;\phi\}\right] \nonumber\\[.5ex]
&g_{u\phi} = -\partial_\phi\omega_1 -e^{-\omega_0}\left[
  \partial_\phi{}^2\left(\frac{\partial_\phi u_0}{\partial_\phi\phi_0}\right) 
   - \frac{\partial_\phi{}^2\phi_0}%
   {\partial_\phi\phi_0}\partial_\phi \left(\frac{\partial_\phi u_0}{\partial_\phi\phi_0}\right)
  + (\partial_\phi\phi_0)^2\left(\frac{\partial_\phi u_0}{\partial_\phi\phi_0}\right)^2\right]
\label{A2}
\end{align}
The variation (\ref{b4}) of the action also acquires an extra term:
\begin{align}
\delta I = \dots + \frac{1}{\kappa^2}\int_{r={\bar r}}\!d^2x\,\left[
  \delta(e^{\omega_0}g_{uu}^{(0)}) + g_{uu}^{(0)} \delta e^{\omega_0}
  -2\omega_1e^{\omega_0}\partial_u\delta\omega_0\right]
\label{A3}
\end{align}
However, if we define
\begin{align}
{\tilde g}_{uu}^{(0)} = g_{uu}^{(0)} +  2e^{-\omega_0}\partial_u\left(e^{\omega_0}\omega_1\right)
  = g_{uu}^{(0)}\bigr|_{\omega_1=0}
\label{A4}
\end{align}
it is easy to check that
\begin{align}
\delta I = \dots +  \frac{1}{\kappa^2} \int_{r={\bar r}}\!d^2x\,\left[\delta(e^{\omega_0}{\tilde g}_{uu}^{(0)}) 
  + {\tilde g}_{uu}^{(0)} \delta e^{\omega_0}\right]
\label{A5}
\end{align}
exactly as in (\ref{b4}).  The addition on $\omega_1$ thus has no effect on the induced boundary 
action.
\end{appendix}

\vspace{1.5ex}
\begin{flushleft}
\large\bf Acknowledgments
\end{flushleft}

This research was supported by the US Department of Energy under grant DE-FG02-91ER40674.
Portions of the work were performed at the Abdus Salam International Centre for 
Theoretical Physics and at the Perimeter Institute for Theoretical Physics.  Research at 
Perimeter Institute is supported by the Government of Canada through the Department 
of Innovation, Science and Economic Development and by the Province of Ontario through 
the Ministry of Research and Innovation.

\end{document}